# Scale-, shift- and rotation-invariant diffractive optical networks

*Deniz Mengu[1,2,3], Yair Rivenson[1,2,3], Aydogan Ozcan[1,2,3,4,*]*

[1] Electrical and Computer Engineering Department, University of California, Los Angeles, CA, 90095, USA

[2] Bioengineering Department, University of California, Los Angeles, CA, 90095, USA

[3] California NanoSystems Institute, University of California, Los Angeles, CA, 90095, USA

[4] Department of Surgery, David Geffen School of Medicine, University of California, Los Angeles, CA, 90095, USA.

[*]Corresponding author: ozcan@ucla.edu

**ABSTRACT**

Recent research efforts in optical computing have gravitated towards developing optical neural networks that aim to benefit from the processing speed and parallelism of optics/photonics in machine learning applications. Among these endeavors, Diffractive Deep Neural Networks ($D^2NNs$) harness light-matter interaction over a series of trainable surfaces, designed using deep learning, to compute a desired statistical inference task as the light waves propagate from the input plane to the output field-of-view. Although, earlier studies have demonstrated the generalization capability of diffractive optical networks to unseen data, achieving e.g., >98% image classification accuracy for handwritten digits, these previous designs are in general sensitive to the spatial scaling, translation and rotation of the input objects. Here, we demonstrate a new training strategy for diffractive networks that introduces input object translation, rotation and/or scaling during the training phase as uniformly distributed random variables to build resilience in their blind inference performance against such object transformations. This training strategy successfully guides the evolution of the diffractive optical network design towards a solution that is scale-, shift- and rotation-invariant, which is especially important and useful for dynamic machine vision applications in e.g., autonomous cars, in-vivo imaging of biomedical specimen, among others.

**Introduction**

Motivated by the success of deep learning[1,2] in various applications[3–16], optical neural networks have gained an important momentum in recent years. Although optical neural networks and related optical computing hardware are relatively at an earlier stage in terms of their inference and generalization capabilities, when compared to the state-of-the-art electronic deep neural networks and the underlying digital processors, optics/photonics technologies might potentially bring significant advantages for machine learning systems in terms of their power efficiency, parallelism and computational speed[17–30]. Among different physical architectures used for the design of optical neural networks[17,22–25,27,31,32], Diffractive Deep Neural Networks ($D^2NN$s)[31,33–39] utilize the diffraction of light through engineered surfaces/layers to form an optical network that is based on light-matter interaction and free-space propagation of light. $D^2NN$s offer a unique optical machine learning framework that formulates a given learning task as a black-box function approximation problem, parameterized through the trainable physical features of matter that control the phase and/or amplitude of light. One of the most convenient methods to devise a $D^2NN$ is to employ multiple transmissive and/or reflective diffractive surfaces/layers that collectively form an optical network between an input and output field-of-view. During the training stage, the complex-valued transmission/reflection coefficients of the layers of a $D^2NN$ are designed for a given statistical (or deterministic) task/goal, where each diffractive feature (i.e., neuron) of a given layer is iteratively adjusted during the training phase using e.g., the error back-propagation method[30,40,41]. After this training and design phase, the resulting diffractive layers/surfaces are physically fabricated using e.g., 3D printing or lithography, to form a passive optical network that performs inference as the input light diffracts from the input plane to the output. Alternatively, the final diffractive layer models can also be implemented by using various types of spatial light modulators (SLMs) to bring reconfigurability and data adaptability to the diffractive network, at the expense of e.g., increased power consumption of the system.

Since the initial experimental demonstration of image classification using D²NNs that are composed of 3D-printed diffractive layers[31,36], the optical inference capacity of diffractive optical networks has been significantly improved based on e.g., differential detection scheme, class-specific designs and ensemble-learning techniques[33,34]. Owing to these systematic advances in diffractive optical networks and training methods, recent studies have reported classification accuracies of >98%, >90% and >62% for the datasets of handwritten digits (MNIST), fashion products (Fashion-MNIST) and CIFAR-10 images, respectively.[33,34] Beyond classification tasks, diffractive networks were also shown to serve as trainable optical front-ends, forming hybrid (optical-electronic) machine learning systems[35]. Replacing the conventional imaging-optics in machine vision systems with diffractive optical networks has been shown to offer unique opportunities to lower the computational complexity and burden on back-end electronic neural networks as well as to mitigate the inference accuracy loss due to pixel-pitch limited, low-resolution imaging systems.[35] Furthermore, in a recent study, diffractive optical networks have been trained to encode the spatial information of input objects into the power spectrum of the diffracted broadband light, enabling object classification and image reconstruction using only a single-pixel spectroscopic detector at the output plane, demonstrating an unconventional, task-specific and resource-efficient machine vision platform.[36]

In all of these existing diffractive optical network designs, the inference accuracies are in general sensitive to object transformations such as e.g., lateral translation, rotation, and/or scaling of the input objects that are frequently encountered in various machine vision applications. In this work, we quantify the sensitivity of diffractive optical networks to these uncertainties associated with the lateral position, scale and in-plane orientation/rotation angle of the input objects (see Fig. 1). Furthermore, we demonstrate a D²NN design scheme that formulates these object transformations through random variables used during the deep learning-based training phase of the diffractive layers. In this manner, the evolution of the layers of a diffractive optical network can adapt to random translation, scaling and rotation of the input objects and, hence, the blind inference capacity of the optical network can be maintained despite these input object uncertainties. The presented training strategy will enable diffractive optical networks to find applications in machine vision systems that require low-latency as well as memory- and power-efficient inference engines for monitoring dynamic events. Beyond diffractive networks, the outlined training scheme can be utilized in other optical machine learning platforms as well as in deep learning-based inverse design problems to create robust solutions that can sustain their target performance against undesired/uncontrolled input field transformations.

**Results and Discussion**

In a standard $D^2NN$-based optical image classifier[31,33–35,42], the number of opto-electronic detectors positioned at the output plane is equal to the number of classes in the target dataset and, each detector uniquely represents one data class (see Fig. 1a). The final class decision is based on the *max* operation over the collected optical signals by these class detectors. According to the diffractive network layout illustrated in Fig. 1a, the input objects (e.g., handwritten MNIST digits) lie within a pre-defined field-of-view (FOV) of $53.33\lambda \times 53.33\lambda$, where $\lambda$ denotes the wavelength of the illumination light. The center of the FOV coincides with the optical axis passing through the center of the diffractive layers. The size of each diffractive layer is chosen to be $106.66\lambda \times 106.66\lambda$, i.e., exactly 2× the size of the input FOV on each lateral axis. The smallest diffractive feature size on each $D^2NN$ layer is set to be ~$0.53\lambda$, i.e., there are 200×200 trainable features on each diffractive layer of a given $D^2NN$ design. At the output plane, each detector is assumed to cover an area of $6.36\lambda \times 6.36\lambda$ and they are located within an output FOV of $53.33\lambda \times 53.33\lambda$ – matching the input FOV size.

Based on these design parameters, a 5-layer diffractive optical network with phase-only modulation at each neuron achieves a blind testing accuracy of 97.64% for the classification of amplitude-encoded MNIST images illuminated with a uniform plane wave. Figure 2a illustrates the thickness profiles of the resulting 5 diffractive layers, constituting this standard D²NN design. To quantify the sensitivity of the blind inference accuracy of this D²NN design against uncontrolled lateral object translations, we introduced an object displacement vector (Fig. 1b), $\mathbf{D} = (D_x, D_y)$, that has two components, defined as independent, uniformly distributed random variables:

$$D_x \sim U(-\Delta_x, \Delta_x) \qquad \text{1a,}$$

$$D_y \sim U(-\Delta_y, \Delta_y) \qquad \text{1b.}$$

The standard diffractive network model (shown in Fig. 2a) was trained (*tr*) with $\Delta_x = \Delta_y = \Delta_{tr} = 0$, and was then tested under different levels of input object position shifts by sweeping the values of $\Delta_x = \Delta_y = \Delta_{test}$ from 0 to 33.92λ with steps of 0.53λ. Stated differently, the final test accuracy corresponding to each $\Delta_{test}$ value reflects the image classification performance of the same diffractive network model that was tested with 10,000 different object positions randomly chosen within the range set by $\Delta_{test}$ (see Fig. 3a for exemplary test objects). This analysis revealed that the blind inference accuracy of the standard D²NN design ($\Delta_{tr} = 0$) which achieves 97.64% under $\Delta_{test} = 0$ quickly falls below 90% as the input objects starts to move within the range $\mp 3.5\lambda$ (blue curve in Fig. 3, defined with $\Delta_{tr} = 0$). As the area covered by the possible object shifts is increased further, the inference accuracy of this native network model decreases rapidly (see Fig. 3).

In this conventional design approach, the optical forward model of the diffractive network training assumes that the input objects inside the sample FOV are free-from any type of undesired geometrical variations, i.e., $\Delta_{tr}= 0$. Hence, the diffractive layers are *not* challenged to process optical waves coming from input objects at different spatial locations, possibly overfitting to the assumed FOV location. As a result, the inference performance of the resulting diffractive network model becomes dependent on the relative lateral location of the input object with respect to the plane of the diffractive layers and the output detectors.

To mitigate this problem, we adopted a training strategy inspired by data augmentation techniques used in deep learning. According to this scheme, each training image sample in a batch is randomly shifted, based on a realization of the displacement vector (***D***), and subsequently, the loss function is computed by propagating these randomly shifted object fields through the diffractive network (see the Methods for details). Using this training scheme, we designed 5 different diffractive network models based on different ranges of object displacement, i.e., $\Delta_x= \Delta_y= \Delta_{tr}= 2.12\lambda, 4.24\lambda, 8.48\lambda, 16.96\lambda \text{ and } 33.92\lambda$ (see Eq. 1). Figure 3 illustrates the MNIST image classification accuracies provided by these 5 new diffractive network models as a function of $\Delta_{test}$. Comparison between the diffractive network models trained with $\Delta_{tr}= 0$ (blue) and $\Delta_{tr}= 2.12\lambda$ (red) reveals that due to the data augmentation introduced by the small object shifts during the training, the latter can achieve an improved inference accuracy of 98.00% for MNIST digits under $\Delta_{test}= 0$. Furthermore, the diffractive network trained with $\Delta_{tr}= 2.12\lambda$ can maintain its classification performance when the input objects are randomly shifted within a certain lateral range (see the right shift of the red curve in Fig. 3). Similarly, training a diffractive network model with $\Delta_{tr}= 4.24\lambda$ (yellow curve in Fig. 3) also results in a better classification accuracy of 97.75% when compared to the 97.64% achieved by the standard model ($\Delta_{tr}= 0$) under $\Delta_{test}= 0$. In addition, this new diffractive model exhibits further resilience to random shifts of the objects within the input FOV, which is indicated by the stronger right shift of the yellow curve in Fig. 3.

For example, for $\Delta_{test}= 3.71\lambda$ in Fig. 3, the input test objects are randomly shifted in *x* and *y* by an amount determined by $D_x \sim U(-3.71\lambda, 3.71\lambda)$ and $D_y \sim U(-3.71\lambda, 3.71\lambda)$, respectively, and this results in a classification accuracy of 97.07% for the new diffractive model ($\Delta_{tr}= 4.24\lambda$), whereas the inference accuracy of the standard model ($\Delta_{tr}= 0$) decreases to 89.88% under the same random lateral shifts of the input test objects.

Further increasing the range of the object location uncertainty, e.g., to $\Delta_{tr}= 8.48\lambda$ (purple curve in Fig. 3), we start to observe a trade-off between the peak inference accuracy and the resilience of the diffractive network to random object shifts. For instance, the diffractive optical network trained with $\Delta_{tr}= 8.48\lambda$ can achieve a peak classification accuracy of 95.55%, which represents a ~2% accuracy compromise with respect to the native diffractive network model ($\Delta_{tr}= 0$) tested under $\Delta_{test}= 0$. However, using such a large object location uncertainty in the training phase also results in a rather flat accuracy curve over a much larger $\Delta_{test}$ range as shown in Fig. 3; in other words, this design strategy expands the effective input object FOV that can be utilized for the desired machine learning task. For example, if the test objects were to freely move within the area defined by $\Delta_x= \Delta_y= \Delta_{test}= 6.89\lambda$, the diffractive network model trained with $\Delta_{tr}= 8.48\lambda$ (purple curve in Fig. 3) brings a >30% inference accuracy advantage compared to the standard model (blue curve in Fig. 3). The resulting layer thickness profiles for this diffractive optical network design trained with $\Delta_{tr}= 8.48\lambda$ are also shown in Fig. 2b.

For the case where $\Delta_{tr}$ was set to be 16.96$\lambda$, the mean test classification accuracy over the range $0 < \Delta_{test} < \Delta_{tr}$ is observed to be 90.46% (see the green curve in Fig. 3b). The relatively more pronounced performance trade-off in this case can be explained based on the increased input FOV. Stated differently, with larger $\Delta_{tr}$ values, the effective input FOV of the diffractive network is increased, and the dimensionality of the solution space[37] provided by a diffractive network design with a limited number of layers (and neurons) might not be sufficient to provide the desired solution when compared to a smaller input FOV diffractive network design. The use of wider diffractive layers (i.e., larger number of neurons per layer) can be a strategy to further boost the inference accuracy over larger $\Delta_{tr}$ values (or larger effective input FOVs), which will be further discussed and demonstrated later in our analysis below (see Fig. 4b).

As an alternative design strategy, the detector plane configuration shown in Fig. 1a can also be replaced with a differential detection scheme[33] to mitigate this relative drop in blind inference accuracy for designs with large $\Delta_{tr}$. In this scheme, instead of assigning a single optoelectronic detector per class, we designate two detectors to each data class and represent the corresponding class scores based on the normalized difference between the optical signals collected by each detector pair. Figure 4a illustrates a comparison between the blind classification accuracies of standard (solid curves) and differential (dashed curves) diffractive network designs, when they were trained with random lateral shifts of the input objects. For all of these designs, except the $\Delta_{tr}= 33.92\lambda$ case, the differential diffractive networks achieve higher classification accuracies throughout the entire testing range, showing their superior robustness and adaptability to input field variations compared to their non-differential counterparts. For example, the peak inference accuracy (95.55%) achieved by the diffractive optical network trained with $\Delta_{tr}= 8.48\lambda$ (solid purple curve in Fig. 4a) increases to 97.33% using the differential detection scheme (dashed purple curve in Fig. 4a). As another example, for $\Delta_{tr}= 16.96\lambda$, the mean classification accuracy of the differential diffractive network over $0 < \Delta_{test} < \Delta_{tr}$ yields 93.38%, which is ~3% higher compared to the performance of its non-differential counterpart for the same test range.

On the other hand, enlarging the uncertainty in the input object translation further, e.g., $\Delta_{tr}$ = 33.92$\lambda$, starts to balance out the benefits of using differential detection at the output plane (see the solid and dashed blue curves in Fig. 4, which closely follow each other). In fact, when $\Delta_x$ and $\Delta_y$ in Eq. 1 are large enough, such as $\Delta_{tr}$ = 33.92$\lambda$, the effective input FOV increases considerably with respect to the size of the diffractive layers; as we discussed earlier, the use of wider diffractive layers with larger numbers of neurons per layer could be used to mitigate this and improve inference performance of D²NN designs that are trained with relatively large $\Delta_{tr}$ values. To shed more light on this, using $\Delta_{tr}$ = 33.92$\lambda$ we trained two additional diffractive optical network models with wider diffractive layers that cover m=4 and m=9 fold larger number of neurons per layer compared to the standard design (m=1) that has 40K neurons per diffractive layer; stated differently, each diffractive layer of these two new designs contain (2×200)×(2×200) = 4×40K and (3×200)×(3×200) = 9×40K neurons per layer, covering 5 diffractive layers, same as the standard D²NN design. The comparison of the blind classification accuracies of these 5-layer D²NN designs with m=1, 4 and 9, all trained with $\Delta_{tr}$ = 33.92$\lambda$, reveals that an increase in the width of the diffractive layers not only increases the input numerical aperture (NA) of the diffractive network, but also significantly improves the classification accuracies even under large $\Delta_{test}$ (see Fig. 4b). For example, the D²NN design with $\Delta_{tr}$ = 33.92$\lambda$ and m=4 achieves classification accuracies of 83.08% and 85.76% for the testing conditions, $\Delta_{test}$ = 0.0$\lambda$ and $\Delta_{test}$ = $\Delta_{tr}$ = 33.92$\lambda$, respectively. With the same $\Delta_{test}$ values, the diffractive network with m=1, i.e., 40K neurons per layer can only achieve classification accuracies of 79.23% and 81.98%, respectively. The expansion of the diffractive layers to accommodate 9×40K neurons per layer (m=9), further increases the mean classification accuracies over the entire $\Delta_{test}$ range, as illustrated in Fig. 4b.

Next, we expanded the presented training approach to design diffractive optical network models that are resilient to the *scale* of the input objects. To this end, similar to Eqs. 1a and 1b, we defined a scaling parameter, $K \sim U(1-\zeta, 1+\zeta)$, randomly covering the scale range $(1-\zeta, 1+\zeta)$ determined by the hyperparameter, $\zeta$. According to this formulation, for a given value of $K$, the physical size of the input object is scaled up ($K > 1$) or down ($K < 1$); see Fig. 5a. Based on this formulation, in addition to the standard D²NN design with $\zeta_{tr} = 0$, we trained 4 new diffractive network models with $\zeta_{tr} = 0.1, 0.2, 0.4$ and $0.8$. The resulting diffractive network models were then tested by sweeping $\zeta_{test}$ from 0 to 0.8 with steps of 0.02 and for each case, the classification accuracy on testing data attained by each diffractive model was computed (see Fig. 5b). This analysis reveals that the resulting diffractive network designs are rather resilient to random scaling of the input objects, maintaining a competitive inference performance over a large range of object shrinkage or expansion (Fig. 5b). Similar to the case shown in Fig. 3, the relatively small values of $\zeta_{tr}$, e.g., 0.1 (red curve in Fig. 5b) or 0.2 (yellow curve in Fig. 5b), effectively serve as data augmentation and the corresponding diffractive network models achieve higher peak inference accuracies of 97.84% ($\zeta_{tr} = 0.1$) and 97.88% ($\zeta_{tr} = 0.2$) compared to the 97.64% achieved by the standard design ($\zeta_{tr} = 0$). Furthermore, the comparison between the shift- and scale-invariant diffractive optical network models trained with $\Delta_{tr} = 16.96\lambda$ (green curve in Fig. 3b) and $\zeta_{tr} = 0.8$ (green curve in Fig. 5b) is highly interesting since the effective FOVs induced by these two training parameters at the input/object plane are quite comparable, resulting in ~1.87× and 1.8× of the FOV of the standard design ($\Delta_{tr} = \zeta_{tr} = 0$), respectively. Despite these comparable effective FOVs at the input plane, the diffractive network trained against random scaling, $\zeta_{tr} = 0.8$, achieves nearly ~6% higher inference accuracy compared to the shift-invariant design, $\Delta_{tr} = 16.96\lambda$. The mean classification accuracy provided by this scale-invariant diffractive optical network model ($\zeta_{tr} = 0.8$) over the entire testing range, $0 < \zeta_{test} < 0.8$, is found to be 96.57% (Fig. 5b), which is only ~1% lower than that of the standard diffractive design tested in the absence of random object scaling ($\zeta_{test} = 0$).

To explore if there is a large performance gap between the classification accuracies attained for de-magnified and magnified input objects, next we *separately* tested the diffractive optical network models in Fig. 5b for the case of expansion-only, i.e., $K \sim U(1, 1 + \zeta)$ and shrinkage-only, i.e., $K \sim U(1 - \zeta, 1)$; see Fig. 5c. A comparison of the solid (expansion-only) and the dashed (shrinkage-only) curves in Fig. 5c reveals that, in general, diffractive networks' resilience toward object expansion and object shrinkage is similar. For instance, for the case of $\zeta_{tr} = 0.4$ (purple curves in Fig. 5c) the mean classification accuracy difference observed between the expansion-only vs. shrinkage-only testing is only 0.04% up to the point that the testing range is equal to that of the training, i.e., $\zeta_{test} = \zeta_{tr}$. Similarly, for $\zeta_{tr} = 0.8$ the mean classification accuracy difference observed between the expansion-only vs. shrinkage-only testing is ~0.75%. When analyzing these results reported in Fig. 5c, one should carefully consider the fact that for a fixed choice of $\zeta$ parameter there is an inherent asymmetry in expansion and shrinkage percentages; for example, for $\zeta_{test} = 0.8$, $K$ can take values in the range (0.2,1.8), where the extreme cases of 0.2 and 1.8 correspond to 5× shrinkage and 1.8× expansion of the input object, respectively. Therefore, the curves reported in Fig. 5c for expansion-only vs. shrinkage-only testing naturally contain different percentages of scaling with respect to the original size of the input objects.

Next, we expanded the presented framework to handle input object *rotations*. Figure 6 illustrates an equivalent analysis as in Fig. 3, except that the input objects are now rotating, instead of shifting, around the optical axis, according to a uniformly distributed random rotation angle, $\Theta \sim U(-\theta, \theta)$, where $\Theta < 0$ and $\Theta > 0$ correspond to clockwise and counterclockwise rotation as depicted in Fig. 1b, respectively. In this comparative analysis, six different diffractive network models trained with $\theta_{tr}$ values taken as 0º (standard design), 5º, 10º, 20º, 30º and 60º were tested as a function of $\theta_{test}$ taking values between 0º and 60º with a step size of 1º, i.e., $\Theta \sim U(-\theta_{test}, \theta_{test})$. Similar to the case of scale-invariant designs reported in Fig. 5, these diffractive network models trained with different $\theta_{tr}$ values can build up strong resilience against random object rotations, almost without a compromise in their inference. In fact, training with $\theta_{tr} \leq 20º$ (red, yellow and purple curves in Fig. 6b) improves the peak inference accuracy over the standard design ($\theta_{tr} = 0°$). When $\theta_{tr} = 30°$ (green curve in Fig. 6b), the inference of the diffractive optical network is relatively flat as a function of $\theta_{test}$, achieving a classification accuracy of 97.51% and 96.68% for $\theta_{test} = 0º$ and $\theta_{test} = 30º$, respectively, clearly demonstrating the advantages of the presented design framework.

Finally, we investigated the design of diffractive optical network models that were trained to simultaneously accommodate two of the three commonly encountered input objects transformations, i.e., random lateral shifting, scaling and in-plane rotation. Table 1 reports the resulting classification accuracies of these newly trained D²NN models, where the inference performance of the corresponding diffractive optical network was tested with the same level of random object transformation as in the training, i.e., $\Delta_{tr} = \Delta_{test}$, $\zeta_{tr} = \zeta_{test}$, $\theta_{tr} = \theta_{test}$. The results in Table 1 reveal that these diffractive network designs can maintain their inference accuracies over 90%, building up resilience against unwanted, yet practically-inevitable object transformations and variations. The thickness profile of the diffractive layers constituting the D²NN designs trained with the object transformation parameter pairs: ($\Delta_{tr} = 2.12\lambda, \theta_{tr} = 10°$), ($\Delta_{tr} = 2.12\lambda, \zeta_{tr} = 0.4$) and ($\theta_{tr} = 10°, \zeta_{tr} = 0.4$) reported in Table 1 are illustrated in Fig. S1 of the Supporting Information. The confusion matrices provided by these three diffractive network models computed under $\Delta_{tr} = \Delta_{test}$, $\zeta_{tr} = \zeta_{test}$, and $\theta_{tr} = \theta_{test}$, are also reported in Fig. S2.

**Conclusions**

In conclusion, we have quantified the sensitivity of diffractive optical networks against three fundamental object transformations (lateral translation, scaling and rotation), that are frequently encountered in various machine vision applications. Moreover, a new design scheme that formulates these input field transformations through uniformly distributed random variables as part of the optical forward model has been presented in deep learning-based training of D$^2$NNs. Our analyses reveal that this training strategy significantly increases the robustness of diffractive networks against undesired object field transformations. Although, we have taken input object classification as our target inference task, the presented ideas and the underlying methods can be extended to other optical machine learning tasks. As the presented training scheme enables the diffractive optical networks to achieve significantly higher inference accuracies in dynamic environments, we believe that this study will potentially expand the utilization of diffractive networks to a plethora of new applications that demand real-time monitoring and classification of fast and dynamic events.

**Methods**

D$^2$NN framework formulates the all-optical object classification problem from the point-of-view of training the physical features of matter inside a diffractive optical black-box. In this study, we modeled each D²NN using 5 successive modulation layers, each representing a two-dimensional, thin modulation component (Fig. 1a). The optical modulation function of each diffractive layer was sampled with a period of 0.53λ over a regular 2D grid of coordinates, with each point representing the transmittance coefficient of a diffractive feature, i.e., an optical "neuron". Following earlier work[36,38,39,42], we selected the material thickness, $h$, as the trainable physical parameter of each neuron,

$$h = Q_4(\frac{\sin(h_a) + 1}{2}(h_m - h_b)) + h_b \qquad 2,$$

According to Eq. 2, the material thickness over each diffractive neuron is defined as a function of an auxiliary variable, $h_a$. The function, $Q_n(.)$, represents the n-bit quantization operator and $h_m$, $h_b$ denote the pre-determined hyperparameters of our forward model determining the allowable range of thickness values, $[h_b, h_m]$. The thickness in Eq. 2 is related to the transmittance coefficient of the corresponding diffractive neuron through the complex-valued refractive index ($\tau$) of the optical material used to fabricate the resulting D²NN, i.e., $\tau(\lambda) = n(\lambda) + j\kappa(\lambda)$, with $\lambda$ denoting the wavelength of the illumination light. Based on this, we can express the transmission coefficient, $t(x_q, y_p, z_k)$, of a diffractive neuron located at $(x_q, y_p, z_k)$ as;

$$t(x_q, y_p, z_k) = \exp\left(-\frac{2\pi\kappa h_{q,p}^k}{\lambda}\right) \exp\left(j(n - n_s)\frac{2\pi h_{q,p}^k}{\lambda}\right) \qquad 3,$$

where $h_{q,p}^k$ refers to the material thickness over the corresponding neuron computed using Eq. 2, and $n_s$ is the refractive index of the medium, surrounding the diffractive layers; without loss of generality, we assumed $n_s = 1$ (air). Based on the earlier demonstrations of diffractive optical networks[31,36,38,39,42], we assumed the optical modulation surfaces in our diffractive optical networks are made of a material with $\tau = 1.7227 + j0.031$. Accordingly, the $h_m$ and $h_b$ were selected as $2\lambda$ and $0.66\lambda$, respectively, as illustrated in Fig. 2 and Fig. S1.

The 2D complex modulation function, $T(x, y, z_k)$, of a diffractive surface, $S_k$, located at $z = z_k$, can be written as:

$$T(x, y, z_k) = \sum_q \sum_p t(x_q, y_p, z_k) P(x - qw_x, y - pw_y, z_k) \qquad 4,$$

where the $w_x$ and $w_y$ denote the width of a diffractive neuron in x and y directions, respectively (both taken as $0.53\lambda$). $P(x, y, z_k)$ represents the 2D interpolation kernel which we assumed to be an ideal rectangular function in the following form,

$$P(x, y, z_k) = \begin{cases} 1, & |x| < \left(\frac{w_x}{2}\right) \text{ and } |y| < \left(\frac{w_y}{2}\right) \\ 0, & \text{otherwise} \end{cases} \qquad 5.$$

The light propagation in the presented diffractive optical networks were modeled based on the digital implementation of the Rayleigh-Sommerfeld diffraction equation, using an impulse response defined as:

$$w(x, y, z) = \frac{z}{r^2} \left(\frac{1}{2\pi r} + \frac{1}{j\lambda}\right) \exp(\frac{j2\pi r}{\lambda}) \qquad 6,$$

where $r = \sqrt{x^2 + y^2 + z^2}$. Based on this, the wave field synthesized over a surface at $z = z_{k+1}$, $U(x, y, z_{k+1})$, by a trainable diffractive layer, $S_k$, located at $z = z_k$, can expressed as;

$$U(x, y, z_{k+1}) = U'(x, y, z_k) * w(x, y, z_{k+1} - z_k) \qquad 7,$$

where $U'(x, y, z_k) = U(x, y, z_k) T(x, y, z_k)$ is the complex wave field immediately after the diffractive layer, $k$, and $*$ denotes the 2D convolution operation.

Based on the above outlined optical forward model, if we let the complex-valued object transmittance, $T(x, y, z_0)$, over the input FOV be located at a surface defined with $k = 0$, then the complex field and the associated optical intensity distribution at the *output/detector plane* of a 5-layer diffractive optical network architecture shown in Fig. 1a, can be expressed as $U(x, y, z_6)$ and $I = |U(x, y, z_6)|^2$, respectively. In our forward training model, we assumed that each class detector collects an optical signal, $\Gamma_c$, that is computed through the integration of the output intensity, $I$, over the corresponding detector active area (6.4λ×6.4λ per detector). For a given dataset with C classes, the standard D²NN architecture in Fig. 1a employs C detectors at the output plane, each representing a data class; C=10 for MNIST dataset. Accordingly, at each training iteration, after the propagation of the input object to the output plane (based on Eqs. 7 and 8), a vector of optical signals, $\boldsymbol{\Gamma}$, is formed and then normalized to get $\boldsymbol{\Gamma}'$ using the following relationship:

$$\boldsymbol{\Gamma}' = \frac{\boldsymbol{\Gamma}}{\max\{\boldsymbol{\Gamma}\}} \times T_s \qquad 8,$$

where $T_s$ is a constant temperature parameter[43,44]. Next, the class score of the $c^{th}$ data class, $\sigma_c$, is computed as:

$$\sigma_c = \frac{\exp(\Gamma'_c)}{\sum_{c \in C} \exp(\Gamma'_c)} \qquad 9.$$

In Eq. 10, $\Gamma'_c$ denotes the normalized optical signal collected by the detector, $c$, computed as in Eq. 9. At the final step, the classification loss function, $\mathcal{L}$, in the form of the cross-entropy loss defined in Eq. 11 is computed for the subsequent error-backpropagation and update of the diffractive layers:

$$\mathcal{L} = -\sum_{c \in C} g_c \log(\sigma_c) \quad \quad 10,$$

where ***g*** denotes the one-hot ground truth label vector.

For the digital implementation of the diffractive optical network training outlined above, we developed a custom-written code in Python (v3.6.5) and TensorFlow (v1.15.0, Google Inc.). The backpropagation updates were calculated using the Adam[45] optimizer with its parameters set to be the default values as defined by TensorFlow and kept identical in each model. The learning rate was set to be 0.001 for all the diffractive network models presented here. The training batch sizes were taken as 50 and 20 for the diffractive network designs with 40K neurons per layer and wider diffractive networks reported in Fig. 4b, respectively. The training of a 5-layer diffractive optical network with 40K diffractive neurons per layer takes ~6 hours using a computer with a GeForce GTX 1080 Ti Graphical Processing Unit (GPU, Nvidia Inc.) and Intel® Core ™ i7-8700 Central Processing Unit (CPU, Intel Inc.) with 64 GB of RAM, running Windows 10 operating system (Microsoft). The training of a wider diffractive network presented in Fig. 4b, on the other hand, takes ~30 hours based on the same system configuration due to the larger light propagation windows used in the forward optical model. Since the investigated object transformations were implemented through a custom-developed bilinear interpolation code written based on TensorFlow functions, it only takes ~50 sec longer to complete an epoch with the presented scheme compared to the standard training of $D^2NNs$.

**Acknowledgement**

The Ozcan Research Group at UCLA acknowledges the support of Fujikura (Japan).

**References**

(1) Goodfellow, I.; Bengio, Y.; Courville, A. *Deep Learning*; MIT Press, 2016.
(2) LeCun, Y.; Bengio, Y.; Hinton, G. Deep Learning. *Nature* **2015**, *521* (7553), 436–444. https://doi.org/10.1038/nature14539.
(3) Rivenson, Y.; Wang, H.; Wei, Z.; de Haan, K.; Zhang, Y.; Wu, Y.; Günaydın, H.; Zuckerman, J. E.; Chong, T.; Sisk, A. E.; Westbrook, L. M.; Wallace, W. D.; Ozcan, A. Virtual Histological Staining of Unlabelled Tissue-Autofluorescence Images via Deep Learning. *Nat Biomed Eng* **2019**, *3* (6), 466–477. https://doi.org/10.1038/s41551-019-0362-y.
(4) Wu, Y.; Rivenson, Y.; Wang, H.; Luo, Y.; Ben-David, E.; Bentolila, L. A.; Pritz, C.; Ozcan, A. Three-Dimensional Virtual Refocusing of Fluorescence Microscopy Images Using Deep Learning. *Nature Methods* **2019**, *16* (12), 1323–1331. https://doi.org/10.1038/s41592-019-0622-5.
(5) Rivenson, Y.; Wu, Y.; Ozcan, A. Deep Learning in Holography and Coherent Imaging. *Light: Science & Applications* **2019**, *8* (1), 1–8. https://doi.org/10.1038/s41377-019-0196-0.
(6) Wu, Y.; Luo, Y.; Chaudhari, G.; Rivenson, Y.; Calis, A.; de Haan, K.; Ozcan, A. Bright-Field Holography: Cross-Modality Deep Learning Enables Snapshot 3D Imaging with Bright-Field Contrast Using a Single Hologram. *Light: Science & Applications* **2019**, *8* (1). https://doi.org/10.1038/s41377-019-0139-9.
(7) Amodei, D.; Ananthanarayanan, S.; Anubhai, R.; Bai, J.; Battenberg, E.; Case, C.; Casper, J.; Catanzaro, B.; Cheng, Q.; Chen, G.; Chen, J.; Chen, J.; Chen, Z.; Chrzanowski, M.; Coates, A.; Diamos, G.; Ding, K.; Du, N.; Elsen, E.; Engel, J.; Fang, W.; Fan, L.; Fougner, C.; Gao, L.; Gong, C.; Hannun, A.; Han, T.; Johannes, L. V.; Jiang, B.; Ju, C.; Jun, B.; LeGresley, P.; Lin, L.; Liu, J.; Liu, Y.; Li, W.; Li, X.; Ma, D.; Narang, S.; Ng, A.; Ozair, S.; Peng, Y.; Prenger, R.; Qian, S.; Quan, Z.; Raiman, J.; Rao, V.; Satheesh, S.; Seetapun, D.; Sengupta, S.; Srinet, K.; Sriram, A.; Tang, H.; Tang, L.; Wang, C.; Wang, J.; Wang, K.; Wang, Y.; Wang, Z.; Wang, Z.; Wu, S.; Wei, L.; Xiao, B.; Xie, W.; Xie, Y.; Yogatama, D.; Yuan, B.; Zhan, J.; Zhu, Z. Deep Speech 2 : End-to-End Speech Recognition in English and Mandarin. 10.
(8) Liu, D.; Tan, Y.; Khoram, E.; Yu, Z. Training Deep Neural Networks for the Inverse Design of Nanophotonic Structures. *ACS Photonics* **2018**, *5* (4), 1365–1369. https://doi.org/10.1021/acsphotonics.7b01377.
(9) Rahmani, B.; Loterie, D.; Konstantinou, G.; Psaltis, D.; Moser, C. Multimode Optical Fiber Transmission with a Deep Learning Network. *Light Sci Appl* **2018**, *7* (1), 1–11. https://doi.org/10.1038/s41377-018-0074-1.
(10) Ma, W.; Cheng, F.; Liu, Y. Deep-Learning-Enabled On-Demand Design of Chiral Metamaterials. *ACS Nano* **2018**, *12* (6), 6326–6334. https://doi.org/10.1021/acsnano.8b03569.
(11) Malkiel, I.; Mrejen, M.; Nagler, A.; Arieli, U.; Wolf, L.; Suchowski, H. Plasmonic Nanostructure Design and Characterization via Deep Learning. *Light Sci Appl* **2018**, *7* (1), 60. https://doi.org/10.1038/s41377-018-0060-7.
(12) Wu, Y.; Rivenson, Y.; Zhang, Y.; Wei, Z.; Günaydin, H.; Lin, X.; Ozcan, A. Extended Depth-of-Field in Holographic Imaging Using Deep-Learning-Based Autofocusing and Phase Recovery. *Optica* **2018**, *5* (6), 704. https://doi.org/10.1364/OPTICA.5.000704.

(13) Li, Y.; Xue, Y.; Tian, L. Deep Speckle Correlation: A Deep Learning Approach toward Scalable Imaging through Scattering Media. *Optica, OPTICA* **2018**, *5* (10), 1181–1190. https://doi.org/10.1364/OPTICA.5.001181.

(14) Rivenson, Y.; Liu, T.; Wei, Z.; Zhang, Y.; de Haan, K.; Ozcan, A. PhaseStain: The Digital Staining of Label-Free Quantitative Phase Microscopy Images Using Deep Learning. *Light Sci Appl* **2019**, *8* (1), 23. https://doi.org/10.1038/s41377-019-0129-y.

(15) Chen, L.-C.; Papandreou, G.; Kokkinos, I.; Murphy, K.; Yuille, A. L. DeepLab: Semantic Image Segmentation with Deep Convolutional Nets, Atrous Convolution, and Fully Connected CRFs. *IEEE Transactions on Pattern Analysis and Machine Intelligence* **2018**, *40* (4), 834–848. https://doi.org/10.1109/TPAMI.2017.2699184.

(16) Zhang, Y.; de Haan, K.; Rivenson, Y.; Li, J.; Delis, A.; Ozcan, A. Digital Synthesis of Histological Stains Using Micro-Structured and Multiplexed Virtual Staining of Label-Free Tissue. *Light: Science & Applications* **2020**, *9* (1). https://doi.org/10.1038/s41377-020-0315-y.

(17) Feldmann, J.; Youngblood, N.; Wright, C. D.; Bhaskaran, H.; Pernice, W. H. P. All-Optical Spiking Neurosynaptic Networks with Self-Learning Capabilities. *Nature* **2019**, *569* (7755), 208–214. https://doi.org/10.1038/s41586-019-1157-8.

(18) Hughes, T. W.; Williamson, I. A. D.; Minkov, M.; Fan, S. Wave Physics as an Analog Recurrent Neural Network. *Science Advances* **2019**, *5* (12), eaay6946. https://doi.org/10.1126/sciadv.aay6946.

(19) Shastri, B. J.; Tait, A. N.; Ferreira de Lima, T.; Nahmias, M. A.; Peng, H.-T.; Prucnal, P. R. Neuromorphic Photonics, Principles Of. In *Encyclopedia of Complexity and Systems Science*; Meyers, R. A., Ed.; Springer Berlin Heidelberg: Berlin, Heidelberg, 2018; pp 1–37. https://doi.org/10.1007/978-3-642-27737-5_702-1.

(20) Sande, G. V. der; Brunner, D.; Soriano, M. C. Advances in Photonic Reservoir Computing. *Nanophotonics* **2017**, *6* (3), 561–576. https://doi.org/10.1515/nanoph-2016-0132.

(21) Cichos, F.; Gustavsson, K.; Mehlig, B.; Volpe, G. Machine Learning for Active Matter. *Nature Machine Intelligence* **2020**, *2* (2), 94–103. https://doi.org/10.1038/s42256-020-0146-9.

(22) Bueno, J.; Maktoobi, S.; Froehly, L.; Fischer, I.; Jacquot, M.; Larger, L.; Brunner, D. Reinforcement Learning in a Large-Scale Photonic Recurrent Neural Network. *Optica* **2018**, *5* (6), 756. https://doi.org/10.1364/OPTICA.5.000756.

(23) George, J. K.; Mehrabian, A.; Amin, R.; Meng, J.; de Lima, T. F.; Tait, A. N.; Shastri, B. J.; El-Ghazawi, T.; Prucnal, P. R.; Sorger, V. J. Neuromorphic Photonics with Electro-Absorption Modulators. *Optics Express* **2019**, *27* (4), 5181. https://doi.org/10.1364/OE.27.005181.

(24) Shen, Y.; Harris, N. C.; Skirlo, S.; Prabhu, M.; Baehr-Jones, T.; Hochberg, M.; Sun, X.; Zhao, S.; Larochelle, H.; Englund, D.; Soljačić, M. Deep Learning with Coherent Nanophotonic Circuits. *Nature Photon* **2017**, *11* (7), 441–446. https://doi.org/10.1038/nphoton.2017.93.

(25) Tait, A. N.; de Lima, T. F.; Zhou, E.; Wu, A. X.; Nahmias, M. A.; Shastri, B. J.; Prucnal, P. R. Neuromorphic Photonic Networks Using Silicon Photonic Weight Banks. *Scientific Reports* **2017**, *7* (1). https://doi.org/10.1038/s41598-017-07754-z.

(26) George, J.; Amin, R.; Mehrabian, A.; Khurgin, J.; El-Ghazawi, T.; Prucnal, P. R.; Sorger, V. J. Electrooptic Nonlinear Activation Functions for Vector Matrix Multiplications in Optical Neural Networks. In *Advanced Photonics 2018 (BGPP, IPR, NP, NOMA, Sensors, Networks, SPPCom, SOF)*; OSA: Zurich, 2018; p SpW4G.3. https://doi.org/10.1364/SPPCOM.2018.SpW4G.3.
(27) Mehrabian, A.; Al-Kabani, Y.; Sorger, V. J.; El-Ghazawi, T. PCNNA: A Photonic Convolutional Neural Network Accelerator. In *2018 31st IEEE International System-on-Chip Conference (SOCC)*; 2018; pp 169–173. https://doi.org/10.1109/SOCC.2018.8618542.
(28) Miscuglio, M.; Mehrabian, A.; Hu, Z.; Azzam, S. I.; George, J.; Kildishev, A. V.; Pelton, M.; Sorger, V. J. All-Optical Nonlinear Activation Function for Photonic Neural Networks [Invited]. *Optical Materials Express* **2018**, *8* (12), 3851. https://doi.org/10.1364/OME.8.003851.
(29) Estakhri, N. M.; Edwards, B.; Engheta, N. Inverse-Designed Metastructures That Solve Equations. *Science* **2019**, *363* (6433), 1333–1338. https://doi.org/10.1126/science.aaw2498.
(30) Hughes, T. W.; Minkov, M.; Shi, Y.; Fan, S. Training of Photonic Neural Networks through in Situ Backpropagation and Gradient Measurement. *Optica* **2018**, *5* (7), 864. https://doi.org/10.1364/OPTICA.5.000864.
(31) Lin, X.; Rivenson, Y.; Yardimci, N. T.; Veli, M.; Luo, Y.; Jarrahi, M.; Ozcan, A. All-Optical Machine Learning Using Diffractive Deep Neural Networks. *Science* **2018**, *361* (6406), 1004–1008. https://doi.org/10.1126/science.aat8084.
(32) Prabhu, M.; Roques-Carmes, C.; Shen, Y.; Harris, N.; Jing, L.; Carolan, J.; Hamerly, R.; Baehr-Jones, T.; Hochberg, M.; Čeperić, V.; Joannopoulos, J. D.; Englund, D. R.; Soljačić, M. A Recurrent Ising Machine in a Photonic Integrated Circuit. *arXiv:1909.13877 [physics]* **2019**.
(33) Li, J.; Mengu, D.; Luo, Y.; Rivenson, Y.; Ozcan, A. Class-Specific Differential Detection in Diffractive Optical Neural Networks Improves Inference Accuracy. *AP* **2019**, *1* (4), 046001. https://doi.org/10.1117/1.AP.1.4.046001.
(34) Rahman, S. S.; Li, J.; Mengu, D.; Rivenson, Y.; Ozcan, A. Ensemble Learning of Diffractive Optical Networks. *arXiv:2009.06869 [cs, eees, physics]* 22.
(35) Mengu, D.; Luo, Y.; Rivenson, Y.; Ozcan, A. Analysis of Diffractive Optical Neural Networks and Their Integration With Electronic Neural Networks. *IEEE J. Select. Topics Quantum Electron.* **2020**, *26* (1), 1–14. https://doi.org/10.1109/JSTQE.2019.2921376.
(36) Li, J.; Mengu, D.; Yardimci, N. T.; Luo, Y.; Li, X.; Veli, M.; Rivenson, Y.; Jarrahi, M.; Ozcan, A. Machine Vision Using Diffractive Spectral Encoding. *arXiv:2005.11387 [cs, eesss, physics]* **2020**.
(37) Kulce, O.; Mengu, D.; Rivenson, Y.; Ozcan, A. All-Optical Information Processing Capacity of Diffractive Surfaces. *arXiv:2007.12813 [eees, cs, cs, physics]* 28.
(38) Veli, M.; Mengu, D.; Yardimci, N. T.; Luo, Y.; Li, J.; Rivenson, Y.; Jarrahi, M.; Ozcan, A. Terahertz Pulse Shaping Using Diffractive Legos. *arXiv:2006.16599 [cs, physics]*.
(39) Luo, Y.; Mengu, D.; Yardimci, N. T.; Rivenson, Y.; Veli, M.; Jarrahi, M.; Ozcan, A. Design of Task-Specific Optical Systems Using Broadband Diffractive Neural Networks. *Light Sci Appl* **2019**, *8* (1), 112. https://doi.org/10.1038/s41377-019-0223-1.
(40) Psaltis, D.; Brady, D.; Gu, X.-G.; Lin, S. Holography in Artificial Neural Networks. *Nature* **1990**, *343* (6256), 325–330. https://doi.org/10.1038/343325a0.
(41) Wagner, K.; Psaltis, D. Multilayer Optical Learning Networks. *Appl. Opt., AO* **1987**, *26* (23), 5061–5076. https://doi.org/10.1364/AO.26.005061.

(42) Mengu, D.; Zhao, Y.; Yardimci, N. T.; Rivenson, Y.; Jarrahi, M.; Ozcan, A. Misalignment Resilient Diffractive Optical Networks. *Nanophotonics* **2020**, *9* (13), 4207–4219. https://doi.org/10.1515/nanoph-2020-0291.

(43) Guo, C.; Pleiss, G.; Sun, Y.; Weinberger, K. Q. On Calibration of Modern Neural Networks. *JMLR.org* **2017**, *70*.

(44) Laha, A.; Chemmengath, S. A.; Agrawal, P.; Khapra, M.; Sankaranarayanan, K.; Ramaswamy, H. G. On Controllable Sparse Alternatives to Softmax. In *Advances in Neural Information Processing Systems 31*; Bengio, S., Wallach, H., Larochelle, H., Grauman, K., Cesa-Bianchi, N., Garnett, R., Eds.; Curran Associates, Inc., 2018; pp 6422–6432.

(45) Kingma, D. P.; Ba, J. Adam: A Method for Stochastic Optimization. *arXiv:1412.6980 [cs]* **2014**.

**Figures**

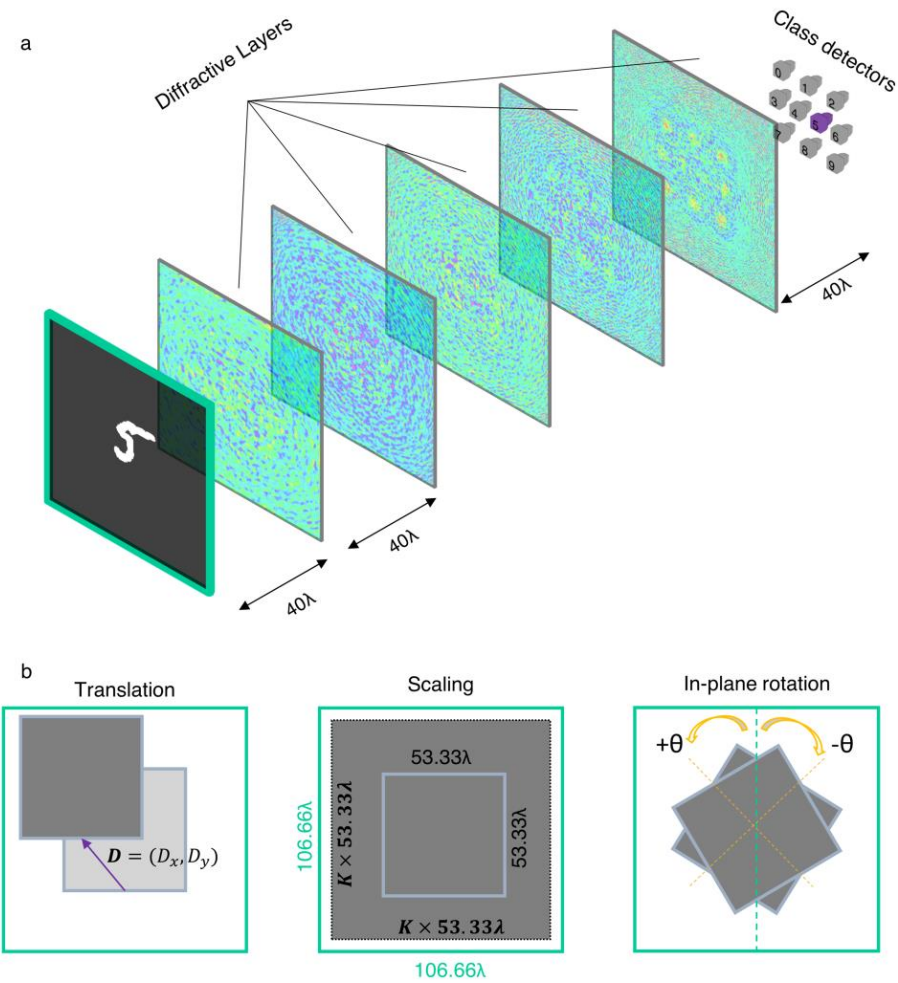

Fig. 1. (a) The layout of the diffractive optical networks trained and tested in this study. (b) The object transformations modeled during the training and testing of the diffractive optical networks presented in this work.

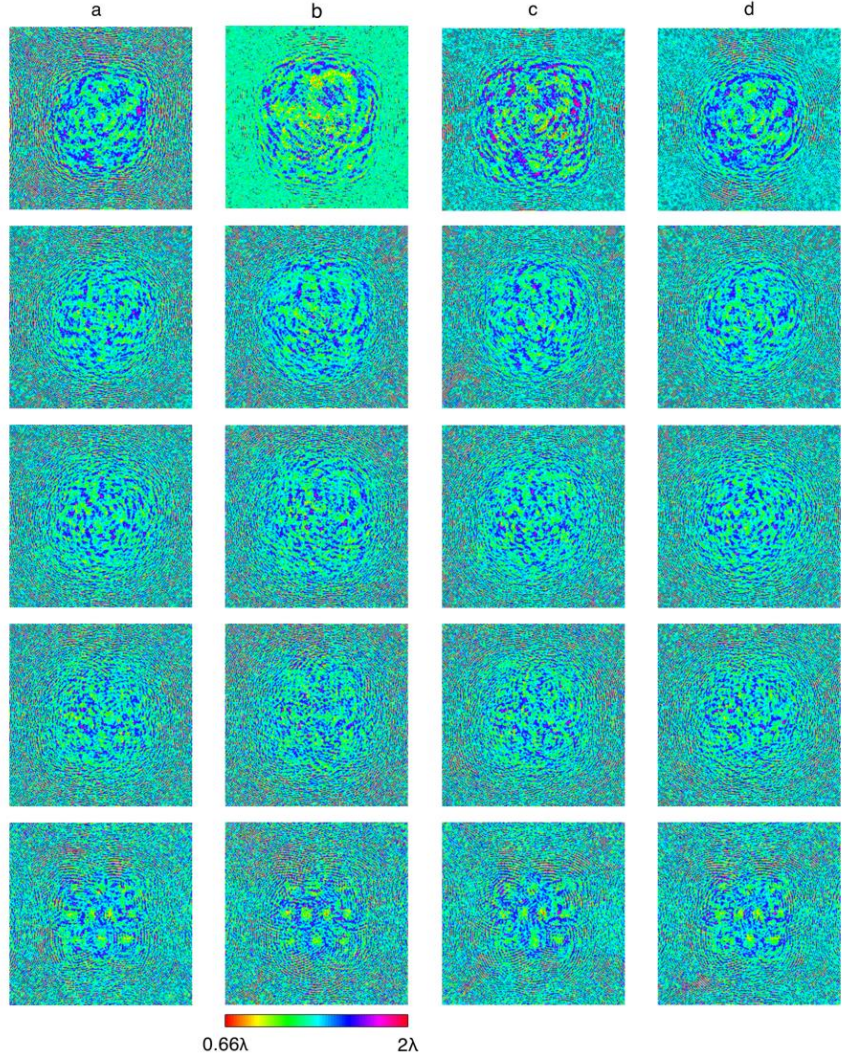

Fig. 2. The thickness profiles of the designed diffractive layers constituting (a) the standard design ($\Delta_{tr} = 0$); (b) the shift-invariant design trained with $\Delta_{tr} = 8.48\lambda$ (purple curve shown in Fig. 3); (c) the scale-invariant design trained with $\zeta_{tr} = 0.4$ (purple curve shown in Fig. 5); (d) the rotation-invariant design trained with $\theta_{tr} = 20°$ (purple curve shown in Fig. 6).

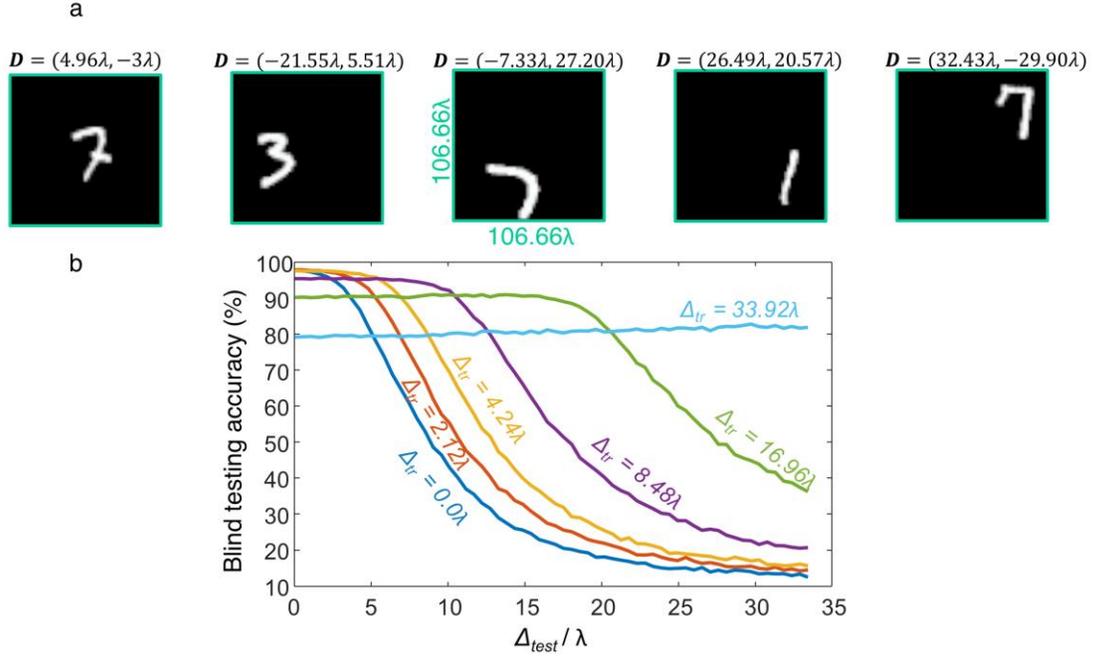

Fig. 3. Shift-invariant diffractive optical networks. (a) Randomly shifted object samples from the MNIST test dataset. Green frame around each object demonstrates the size of the diffractive layers (106.66λ×106.66λ). (b)The blind inference accuracies provided by six different diffractive network models trained with $\Delta_x = \Delta_y = \Delta_{tr}$, taken as 0.0λ (blue), 2.12λ (red), 4.24λ (yellow), 8.48λ (purple), 16.96λ (green), 33.92λ (light-blue) when they were tested under different levels random object shifts with the control parameter, $\Delta_x = \Delta_y = \Delta_{test}$, swept from 0.0λ to 33.92λ.

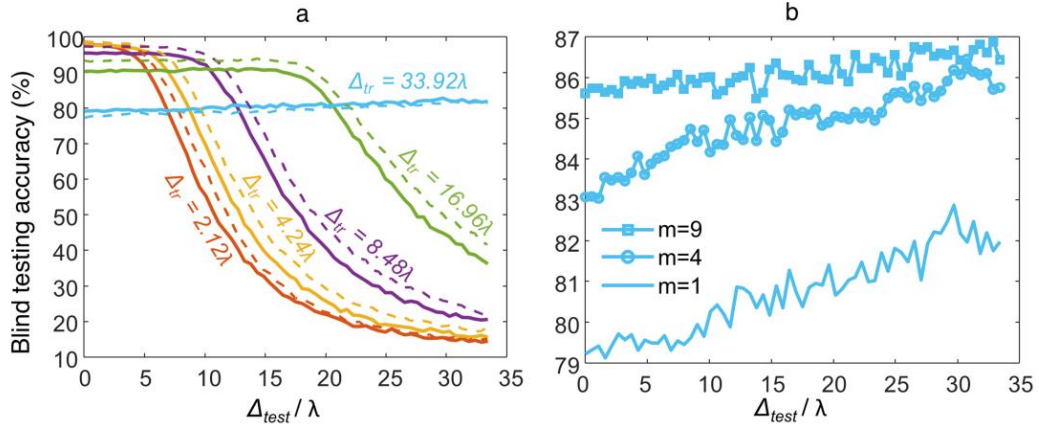

Fig. 4. Different design strategies that can improve the performance of shift-invariant diffractive optical networks. (a) The comparison between the inference accuracies of standard (solid curves) and differential (dashed curves) diffractive optical networks trained using various $\Delta_{tr}$ values. (b) Blind testing classification accuracies of three non-differential, 5-layer D²NN designs that have m×40K optical neurons per layer, with m=1, 4 and 9. All these diffractive optical networks were trained using $\Delta_{tr}$ = 33.92λ. The diffractive network designs with wider diffractive layers and more neurons per layer can generalize more effectively to random object translations.

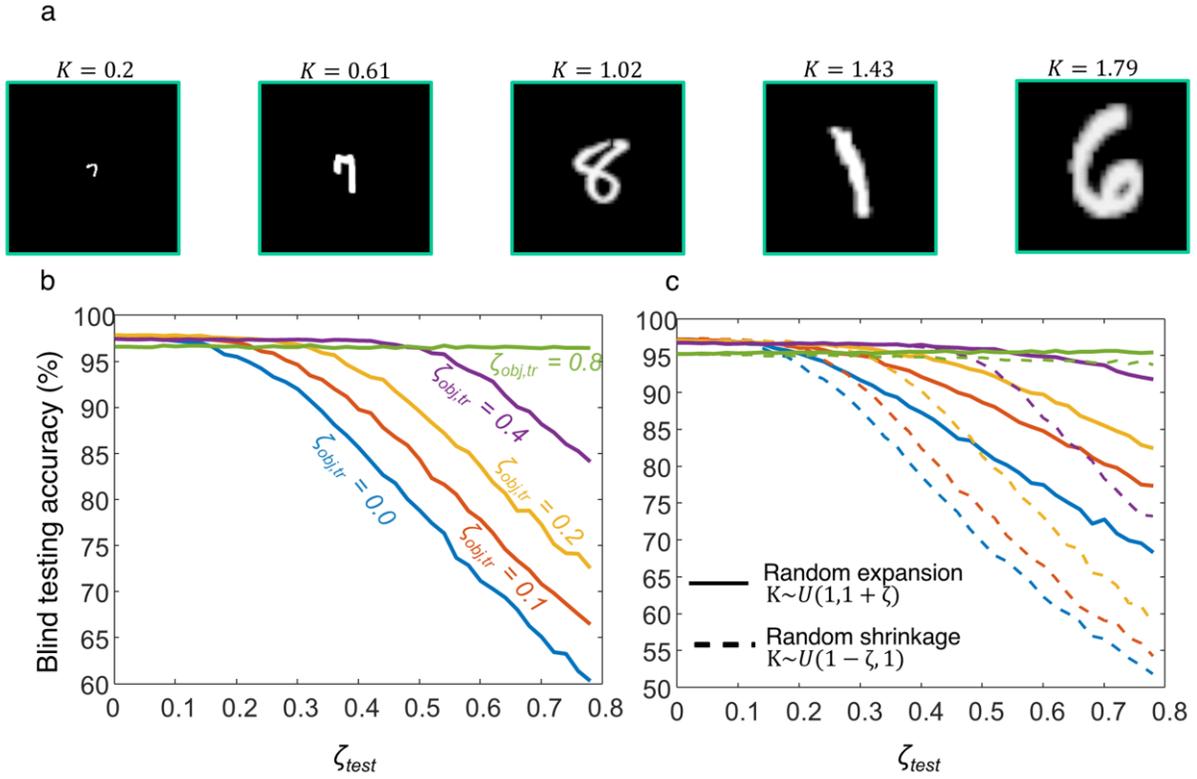

Fig. 5. Scale-invariant diffractive optical networks. (a) Randomly scaled object examples from the MNIST test dataset. Green frame around each object demonstrates the size of the diffractive layers. (b) The blind inference accuracies provided by five different D²NN models trained with $\zeta = \zeta_{tr}$, taken as 0.0 (blue), 0.1 (red), 0.2 (yellow), 0.4 (purple) and 0.8 (green); the resulting models were tested under different levels random object scaling with the parameter, $\zeta = \zeta_{test}$, swept from 0.0 to 0.8. (c) The classification performance of the diffractive networks in (b) for the case of expansion-only (solid curves) and shrinkage-only (dashed curves).

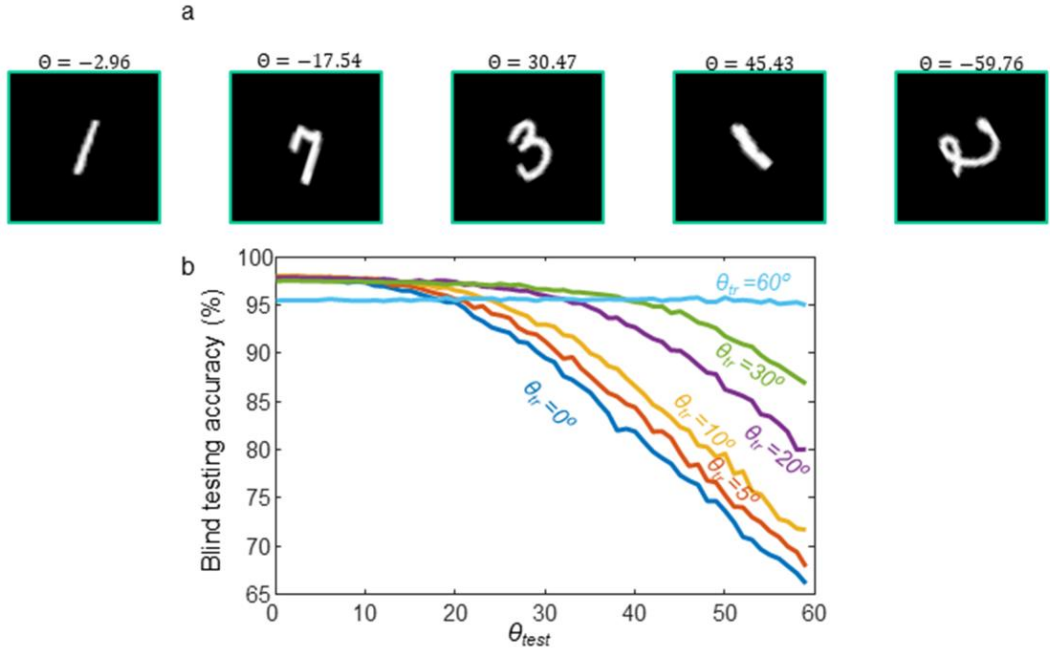

Fig. 6. Rotation-invariant diffractive optical networks. (a) Randomly rotated object examples from the MNIST test dataset. Green frame around each object demonstrates the size of the diffractive layers. (b) The blind inference accuracies provided by five different diffractive network models trained with $\theta = \theta_{tr}$, taken as 0⁰ (blue), 5⁰ (red), 10⁰ (yellow), 20⁰ (purple), 30⁰ (green) and 60⁰ (light-blue) when they were tested under different levels of random object rotations with the parameter, $\theta = \theta_{test}$, swept from 0⁰ to 60⁰, covering both clockwise and counter-clockwise image rotations.

# Tables

## Shift-Rotation

| shift \ rotation | $\Theta_{tr} = \Theta_{test} = 0°$ | $\Theta_{tr}$ and $\Theta_{test}$ are iid $U(-10°, 10°)$ | $\Theta_{tr}$ and $\Theta_{test}$ are iid $U(-20°, 20°)$ |
|---|---|---|---|
| $\Delta_{tr} = \Delta_{test} = 0\lambda$ | 97.64 | 97.64 | 97.13 |
| $\Delta_{tr}$ and $\Delta_{test}$ are iid $U(-2.12\lambda, 2.12\lambda)$ | 97.51 | 97.48 | 96.78 |
| $\Delta_{tr}$ and $\Delta_{test}$ are iid $U(-8.48\lambda, 8.48\lambda)$ | 94.41 | 94.09 | 92.00 |

## Shift-Scaling

| shift \ scaling | $K_{tr} = K_{test} = 1$ | $K_{tr}$ and $K_{test}$ are iid $U(0.6, 1.4)$ | $K_{tr}$ and $K_{test}$ are iid $U(0.2, 1.8)$ |
|---|---|---|---|
| $\Delta_{tr} = \Delta_{test} = 0\lambda$ | 97.64 | 97.29 | 96.50 |
| $\Delta_{tr}$ and $\Delta_{test}$ are iid $U(-2.12\lambda, 2.12\lambda)$ | 97.51 | 96.79 | 95.76 |
| $\Delta_{tr}$ and $\Delta_{test}$ are iid $U(-8.48\lambda, 8.48\lambda)$ | 94.41 | 91.71 | 89.20 |

## Rotation-Scaling

| scaling \ rotation | $\Theta_{tr} = \Theta_{test} = 0°$ | $\Theta_{tr}$ and $\Theta_{test}$ are iid $U(-10°, 10°)$ | $\Theta_{tr}$ and $\Theta_{test}$ are iid $U(-20°, 20°)$ |
|---|---|---|---|
| $K_{tr} = K_{test} = 1$ | 97.64 | 97.64 | 97.13 |
| $K_{tr}$ and $K_{test}$ are iid $U(0.6, 1.4)$ | 97.29 | 96.98 | 96.26 |
| $K_{tr}$ and $K_{test}$ are iid $U(0.2, 1.8)$ | 96.50 | 96.24 | 95.67 |

Table. 1. The blind inference accuracy of the D²NN models trained against the combinations of the three object field transformations investigated in this work: (upper) shift-rotation, (middle) shift-scaling, (lower) rotation-scaling.